\documentclass[twocolumn,showpacs,preprintnumbers,amsmath,amssymb,pra]{revtex4}

\usepackage{array}
\usepackage{graphicx} 
\usepackage{float}
\usepackage{subfigure}
\usepackage{subfigmat}
\usepackage{caption}
\usepackage{dcolumn}
\usepackage{bm}
\usepackage{amsmath}
\usepackage{enumitem}

\newcommand{\ket}[1]{\ensuremath{\left|\left. #1\right.\right>}}
\newcommand{\bra}[1]{\ensuremath{\left<\left.#1\right.\right|}}

\begin{document}


\title{Dissipative Dynamics Of Fermion-Fermion Scattering In A Three-Site Optical Lattice}
\author{Vladimir P. Villegas}
\email{vvillegas@nip.upd.edu.ph}
\author{Roland Cristopher F. Caballar}
\email{rfcaballar@up.edu.ph}
\affiliation{Theoretical Physics Group, National Institute of Physics, University of the Philippines, Diliman Quezon City, 1101 Philippines}

\pacs{03.65.Yz, 05.60.Gg}

\begin{abstract}
In this article, we investigate the dissipative dynamics of a Fermi gas trapped in a three-site optical lattice exposed to a fermionic environment. The lattice sites admit at most one spin-up and one spin-down particle at a time and its interaction with the fermionic environment cause particles to either be trapped by or expelled from it. It is then shown that apart from each lattice site being populated by at least one particle, spin exchange is observed, which allows the possibility for the system to be used to encrypt information via quantum cryptography. Furthermore, we also observe entanglement among the lattice sites, denoting that transport of quantum information among particles is possible in this ultracold atomic system.
\end{abstract}
%
%
\maketitle

\section{Introduction}
\label{sec:intro}

Advances in complex condensed matter systems \cite{martin} and in quantum information \cite{wecker} have stimulated the study of ultracold atomic systems \cite{lewenstein} such as Bose-Einstein condensates (BECs) \cite{caballar, meinert}, Mott insulators \cite{jaksch, bloch1, bloch2}, graphene \cite{schuler}, superconductors \cite{fazio}, and others. Among these systems too are Fermi gases \cite{schreiber}, which is a large ensemble of particles that obey Fermi-Dirac statistics.

In this study, we will deal with a Fermi gas trapped by an optical lattice, which Wecker et al. \cite{wecker} showed to be useful in quantum computing \cite{jaksch, hugel} by determining the phases diagram and ground state properties. Other applications of Fermi gases in optical lattices are in the cooling of magnetically trapped gas \cite{demarco}; observation of quantized vortices, quenching of moment of inertia, and spin polarization in superfluid helium \cite{giorgini}; entanglement in Luttinger liquids \cite{chen}; double-photo-ionization of molecular hydrogen \cite{tichy}; Landau-Zener tunneling in double-well potential in deep BEC regime \cite{wywang}; topological phase transitions driven by real-next-nearest neighbor hopping \cite{rwang}; fermion condensation quantum phase transition in $YbCu_{5-x}Au_{x}$ \cite{shaginyan}; interaction quantum quenches in one-dimensional model with spin imbalance \cite{riegger}; realization of Bardeen-Cooper-Schrieffer (BCS)-BEC state crossover in regime of resonant interactions \cite{bloch2, ohara}; among other applications.

To engineer optical lattice systems, the Hubbard model is used since only nearest neighbor interactions among lattice sites allow the interaction of the Fermi gas particles in the lattice \cite{jaksch, giorgini}. Furthermore, external field parameters can be varied over time using this model \cite{lewenstein, jaksch}. This description of fermionic systems allows us the study of the possibility of quantum phenomena such as spin-exchange interaction and to design properties like anisotropy and sign by proper choices of optical potentials \cite{jaksch}.

Considering the lattice as an open quantum system, i.e. connecting it to an external environment such as a heat bath or a quantum particle (i.e. harmonic oscillator) bath, gives a more realistic description of the system. This is because the energy of the lattice system is not constant due to its exposure to the environment and the number of particles in the lattice is not constant because particles from the bath can be trapped into the lattice and particles in the lattice can be expelled to the bath \cite{bp}.

Since ultracold atomic gases can be trapped by an optical lattice, and taking into account its interaction with the environment, can information be transported from a particular lattice site to another over time such that ultracold fermions trapped in an optical lattice can be used to engineer information transport systems? This is answerable by studying the dissipative dynamics of the Fermi-Hubbard model, i.e. the effect of connecting the optical lattice containing a trapped Fermi gas to a fermionic bath. The characteristics of this system, in particular its ability to transport information and other quantities, can be explored through time by the particle population of the lattice, the existence of spin exchange through time, the particle occupation of each lattice site, and the entanglement in the lattice.

This article studies a system of a Fermi gas trapped by a three-site optical lattice interacting with a fermionic bath in order to observe phenomena such as spin exchange (which can be observed easily through fermions) and entanglement, which can be used to describe information dynamics in systems that can be mapped onto lattices. The discussion is organized as follows: In Section {\ref{sec:fermihubbard}}, based on discussions from Jaksch and Zoller \cite{jaksch} and Breuer and Petruccione \cite{bp}, the model is described by defining the field operators of the system and the bath. The free and interaction Hamiltonians and the Born-Markov master equation of the system are written. The derived dissipative dynamics are then used in Section {\ref{sec:occprobfid}} to describe the transport of information through the lattice using fidelity, occupation probability, and entanglement. Conclusions, recommendations, and possible future extensions are stated in Section {\ref{sec:conclusion}}.

\section{The Fermi-Hubbard Model}
\label{sec:fermihubbard}

\begin{figure}[t]
	\centering
	\subfigure[Open chain.]{\includegraphics[scale=0.5]{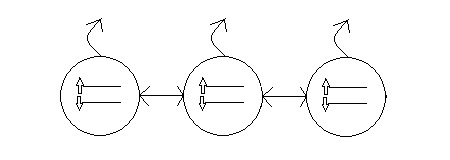}\label{fig:schematicopen}}
	\subfigure[Closed chain.]{\includegraphics[scale=0.5]{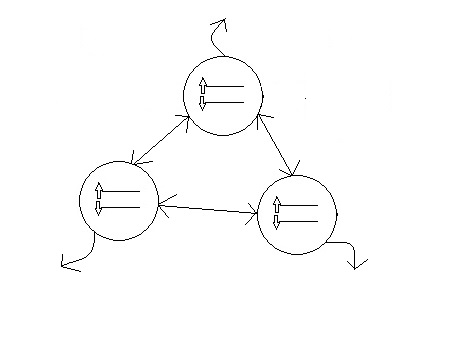}\label{fig:schematicclosed}}
	\caption{Schematic diagram of the system. The optical lattice containing fermionic particles is interacting with a fermionic environment, causing scattering of particles in it. (Environment not shown.)}
	\label{fig:schematic}	
\end{figure}

\subsection{The lattice system}
\label{subsec:lattice system}

A Fermi gas is an ultracold atomic system of particles obeying the Pauli exclusion principle \cite{mahan}. Suppose that the gas is trapped by an optical lattice with a potential $V (\vec{x})$ and whose lattice sites all accommodate just one energy level, with the magnitude of that energy level being identical for all three lattice sites. Since there is only one energy level, each lattice site can be occupied by up to two particles of different spins, i.e. one spin-up and one spin-down. (See Figure \ref{fig:schematic}.)

Mathematically, this system can be represented using the field operator
\begin{equation} \label{latoptr}
	\phi_S = \sum_{\vec{j}, \sigma} \hat{c}_{\vec{j}, \sigma} w_0 (\vec{x} - \vec{x}_{\vec{j}}),
\end{equation}
where $\vec{x}_{\vec{j}}$ is the vector pointing to a specific lattice site and $\sigma \in \{ \uparrow, \downarrow \}$ is the spin of the particle.

The annihilation operator corresponding to the spin of a particle in the lattice site located at $\vec{x}_{\vec{j}}$ is
\begin{equation}
\hat{c}_{\vec{j}, \sigma} = \hat{\sigma}_{-} =
\begin{pmatrix}
0 & 0 \\
1 & 0
\end{pmatrix}.
\label{annmatrix}
\end{equation}
To describe the occupation or non-occupation of a particular spin state, the basis
\begin{eqnarray}
\label{spinbasis}
\ket{0} =
\begin{pmatrix}
0 \\
1
\end{pmatrix}; \;\;
\ket{1} =
\begin{pmatrix}
1 \\
0
\end{pmatrix}
\end{eqnarray}
is used. $\ket{0}$ denotes occupation and $\ket{1}$, non-occupation.

Since a lattice site can be occupied by at most a spin-up and a spin-down particle, the basis for the site's state vectors is $\ket{n_{\uparrow},n_{\downarrow}}_{\vec{j}} = \ket{n_{\uparrow}}_{\vec{j}} \otimes \ket{n_{\downarrow}}_{\vec{j}}$, where $n_{\uparrow},n_{\downarrow} \in \{ 0, 1 \}$ is the number of spin-up and spin-down particles respectively. Therefore, the basis for the state of the particles occupying the lattice site consists of the following column matrices:
\begin{eqnarray}
\label{sitebasis}
\ket{0_{\uparrow}, 0_{\downarrow}}_{\vec{j}} = \ket{0_{\uparrow}}_{\vec{j}} \otimes \ket{0_{\downarrow}}_{\vec{j}} =
\begin{pmatrix}
0 \\
0 \\
0 \\
1
\end{pmatrix}; \\
\ket{0_{\uparrow}, 1_{\downarrow}}_{\vec{j}} = \ket{0_{\uparrow}}_{\vec{j}} \otimes \ket{1_{\downarrow}}_{\vec{j}} =
\begin{pmatrix}
0 \\
0 \\
1 \\
0
\end{pmatrix}; \\
\ket{1_{\uparrow}, 0_{\downarrow}}_{\vec{j}} = \ket{1_{\uparrow}}_{\vec{j}} \otimes \ket{0_{\downarrow}}_{\vec{j}}
\begin{pmatrix}
0 \\
1 \\
0 \\
0
\end{pmatrix}; \\
\ket{1_{\uparrow}, 1_{\downarrow}}_{\vec{j}} = \ket{1_{\uparrow}}_{\vec{j}} \otimes \ket{1_{\downarrow}}_{\vec{j}}
\begin{pmatrix}
1 \\
0 \\
0 \\
0
\end{pmatrix}.
\end{eqnarray}
It then follows that the annihilation operator for the spin-up and the spin-down particles in a lattice site at $\vec{x}_{\vec{j}}$ are
\begin{eqnarray}
	\hat{c}_{\vec{j}, \uparrow} &=& \hat{\sigma}_{-} \otimes \hat{\mathbb{I}} =
	\begin{pmatrix}
	0 & 0 \\
	1 & 0
	\end{pmatrix}
	\otimes
	\begin{pmatrix}
	1 & 0 \\
	0 & 1
	\end{pmatrix}
	, \\
	\hat{c}_{\vec{j}, \downarrow} &=& \hat{\mathbb{I}} \otimes \hat{\sigma}_{-} = \begin{pmatrix}
	1 & 0 \\
	0 & 1
	\end{pmatrix}
	\otimes
	\begin{pmatrix}
	0 & 0 \\
	1 & 0
	\end{pmatrix}
\end{eqnarray}
respectively.

Since the lattice interacts with a bath, it is an open quantum system. Therefore, the state of the system is described using the density matrix
\begin{widetext}
	\label{rhoS}
	\begin{align}
		\rho_S = \sum_{n_{1, \uparrow}, n_{1, \downarrow}, n'_{1, \uparrow}, n'_{1, \downarrow}} \sum_{n_{2, \uparrow}, n_{2, \downarrow}, n'_{2, \uparrow}, n'_{2, \downarrow}} \sum_{n_{3, \uparrow}, n_{3, \downarrow}, n'_{3, \uparrow}, n'_{3, \downarrow}} & \left( \prod_{n; \sigma; \vec{j}} \prod_{n'; \sigma'; \vec{j}'} \alpha_{n; \sigma; \vec{j}} (t) \alpha_{n'; \sigma'; \vec{j}'} (t) \right) \nonumber \\
		& \times \ket{n_{1, \uparrow}, n_{1, \downarrow}}_1 \ket{n_{2, \uparrow}, n_{2, \downarrow}}_2 \ket{n_{3, \uparrow}, n_{3, \downarrow}}_3 \nonumber \\
		& \times \bra{n'_{1, \uparrow}, n'_{1, \downarrow}}_1 \bra{n'_{2, \uparrow}, n'_{2, \downarrow}}_2 \bra{n'_{3, \uparrow}, n'_{3, \downarrow}}_3,
	\end{align}
\end{widetext}
where $0 \leq \alpha_{n; \sigma; \vec{j}} (t) \leq 1$ is the probability of finding $n_{\sigma} \in \{ 0, 1 \}$ particles of certain spin $\sigma$ in a lattice site pointed by $\vec{j}$ at a time $t$.

The potential of the lattice can be split as $V (\vec{x}) = V_T (\vec{x}) + V_0 (\vec{x})$. $V_T (\vec{x})$ is the external trapping potential of the lattice, which varies slowly compared to the solved lattice potential $V_0 (\vec{x})$. An example of trapping potentials are the magnetic trapping potentials, which enable the trapping or expulsion of particles from the optical lattice. Examples of $V_0 (\vec{x})$ are those of particle in a box, harmonic oscillator, and hydrogen atom.

Through Rabi and Raman lasers, we observe that the particle spin splits the known potential such that $V_0 (\vec{x}) = V_0^{Rabi} (\vec{x}) \sigma_{\vec{j}}^z + V_0^{Raman} (\vec{x}) \sigma_{\vec{j}}^x$. This means that the Rabi laser potential $V_0^{Rabi}$ causes the particles to scatter through the lattice with unchanged spin and the Raman laser potential $V_0^{Raman}$ changes the particle spin via scattering.

The free Hamiltonian is defined as
\begin{widetext}
	\begin{align}
	\label{freeintH}
	H_0 = \int d^3 \vec{x} \hat{\phi}_S^\dagger (\vec{x}) \left( \frac{\vec{p}^2}{2 \mu} + V_0 (\vec{x}) + V_T (\vec{x}) \right) \hat{\phi}_S (\vec{x}) + \frac{g}{2} \int d^3 \vec{x} \hat{\phi}_S^\dagger (\vec{x}) \hat{\phi}_S^\dagger (\vec{x}) \hat{\phi}_S (\vec{x}) \hat{\phi}_S (\vec{x}),
	\end{align}
\end{widetext}
where $\mu$ is the reduced mass and $g$ is the interaction strength between two particles. If the only interaction is $s$-wave scattering, $g = 4 \pi a_s / \mu$ where $a_s$ is the $s$-wave scattering length.

This free Hamiltonian can be expressed in terms of the fermionic annihilation and creation operators $\hat{c}_{\vec{j}, \sigma}$ and $\hat{c}_{\vec{j}, \sigma}^{\dagger}$ provided we let the following quantities be constant: the free particle energy
\begin{equation} \label{J}
J_{\vec{j},\vec{k}} \equiv J = - \int d^3 \vec{x} w_{0}^{\ast} (\vec{x} - \vec{x}_{\vec{j}}) \left( \frac{\vec{p}^2}{2 \mu} \right) w_{0} (\vec{x} - \vec{x}_{\vec{k}}) ,
\end{equation}
the energy of interaction between two particles
\begin{equation}
U = g \int w_{0}^{\ast} (\vec{x}-\vec{x}_{\vec{j}}) w_{0}^{\ast} (\vec{x}-\vec{x}_{\vec{k}}) w_{0} (\vec{x}-\vec{x}_{\vec{l}}) w_{0} (\vec{x}-\vec{x}_{\vec{m}}),
\end{equation}
the Rabi laser energy
\begin{equation} \label{rabi}
\delta_R = -2 \int d^3 \vec{x} \phi_S^{\dagger} (\vec{x}) V_0^{sym} (\vec{x}) \phi_S(\vec{x}) ,
\end{equation}
and the Raman laser energy
\begin{equation}
\Omega_R = 2 \int d^3 \vec{x} \phi_S^{\dagger} (\vec{x}) V_0^{ant} (\vec{x}) \phi_S(\vec{x}).
\end{equation}
These quantities are made constant by assumption that they are uniform for all lattice sites in this study.

Thus, in terms of fermionic annihilation and creation operators, the free Hamiltonian is
\begin{widetext}
	\begin{align}
	\label{freeHop}
	H_0 = - J \sum_{\langle \vec{j} \vec{k} \rangle} \sum_{\sigma, \sigma'} \hat{c}_{\vec{j}, \sigma}^{\dagger} \hat{c}_{\vec{k}, \sigma'} + \frac{U}{2} \sum_{\vec{j}, \sigma} \hat{c}_{\vec{j}, \sigma}^{\dagger} \hat{c}_{\vec{j}, \sigma}^{\dagger} \hat{c}_{\vec{j}, \sigma} \hat{c}_{\vec{j}, \sigma} - \frac{\delta_R}{2} \sum_{\vec{j}} (\hat{c}_{\vec{j}, \uparrow}^{\dagger} \hat{c}_{\vec{j}, \uparrow} - \hat{c}_{\vec{j}, \downarrow}^{\dagger} \hat{c}_{\vec{j}, \downarrow}) + \frac{\Omega_R}{2} \sum_{\vec{j}} (\hat{c}_{\vec{j}, \uparrow}^{\dagger} \hat{c}_{\vec{j}, \downarrow} + \hat{c}_{\vec{j}, \downarrow}^{\dagger} \hat{c}_{\vec{j}, \uparrow}).
	\end{align}
\end{widetext}

\subsection{Interaction with a fermionic bath}
\label{subsec:interaction}

We consider the case where the optical lattice interacts with a fermionic bath defined by a set of fermionic annihilation (creation) operators $\hat{h}_{\vec{j}, \sigma}$ ($\hat{h}_{\vec{j}, \sigma}^{\dagger}$). Therefore, the bath is mathematically described using the field operator of the form
\begin{equation} \label{bathfieldop}
\hat{\phi}_B = \sqrt{\rho_C} + \frac{1}{\sqrt{V}} \sum_{\vec{j}} \sum_{\sigma} (u_{\vec{j}} \hat{h}_{\vec{j}, \sigma} e^{i \vec{j} \cdot \vec{x}} + v_{\vec{j}} \hat{h}_{\vec{j}, \sigma}^{\dagger} e^{-i \vec{j} \cdot \vec{x}}),
\end{equation}
where $\rho_C$ is the desnity of the fermionic bath, $V$ is the volume of the bath, $u_{\vec{j}}$ and $v_{\vec{j}}$ are constants.

The state of the bath is described by the density matrix $\rho_B$.

The interaction with the bath causes fermionic particles to either be trapped by or expelled from the lattice. This adds to the effective potential, causing scattering and the spin exchange. Thus, this interaction can be expressed in terms of both the bath and the system field operators such that \cite{caballar}
\begin{equation}
	\label{inthamgen}
	H_I = \frac{2 \pi a_S}{\mu_R} \int d^3 \vec{x} \hat{\phi}_S^{\dagger} \hat{\phi}_S \hat{\phi}_B^{\dagger} \hat{\phi}_B.
\end{equation}

With this study limited to two-body interactions, terms of order $\hat{h}_{\vec{j}, \sigma}^{(\dagger)} \hat{h}_{\vec{j'}, \sigma'}^{(\dagger)}$ can be neglected. Therefore, in terms of annihilation and creation operators of both the system and environment particles, the interaction Hamiltonian of the system is
\begin{widetext}
	\begin{align}
	\label{intHfermi}
	H_I = \frac{2 \pi a_S \rho_C}{\mu_R} \sum_{\vec{l}} \sum_{\sigma, \sigma'} \{ \hat{c}_{\vec{l}, \sigma}^{\dagger} \hat{c}_{\vec{l}, \sigma'} + \frac{1}{2\sqrt{\rho_C V_C}} \sum_{\vec{m}} \sum_{\alpha, \alpha'} A_{\vec{m}, \vec{l}} (u_{\vec{m}} + v_{\vec{m}}) (\hat{c}_{\vec{l}}^{\dagger} \hat{c}_{\vec{l}+ \vec{r}} + h. c. ) \cdot (\hat{h}_{\vec{m}, \alpha} + \hat{h}_{\vec{m'}, \alpha'}^{\dagger}) \}.
	\end{align}
\end{widetext}

See Appendix \ref{app:intham} for the detailed outline of the derivation of the interaction Hamiltonian.

The interaction of the system and the bath evolves through time. Therefore, using the Baker-Campbell-Hausdorff formula, the time-evolved interaction Hamiltonian is
\begin{widetext}
	\begin{align}
	H_I (t) = \frac{2 \pi a \rho_C}{\mu_R} \sum_{ \vec{l} } \sum_{ \alpha, \alpha' } \{ \hat{c}^{\dagger}_{ \vec{l}, \alpha } \hat{c}_{ \vec{l}, \alpha' } + \frac{1}{\sqrt{\rho_C V}} \sum_{ \vec{m} } \sum_{ \beta } & A_{ \vec{m}, \vec{l} } \left( u_{ \vec{m} } + v_{ \vec{m} } \right) \left( \hat{h}_{ \vec{m}, \beta } + \hat{h}^{\dagger}_{ \vec{m}, \beta } \right) \nonumber \\
	& \times \left( \hat{c}^{\dagger}_{ \vec{l}, \uparrow } \hat{c}_{ \vec{l} + \vec{r}, \uparrow } e^{- i \delta_R t} + \left( \hat{c}^{\dagger}_{ \vec{l}, \uparrow } \hat{c}_{ \vec{l} + \vec{r}, \downarrow } + \hat{c}^{\dagger}_{ \vec{l}, \downarrow } \hat{c}_{ \vec{l} + \vec{r}, \uparrow } + h.c. \right) + h.c. \right) \}.
	\label{fermifermiintHt}
	\end{align}
\end{widetext}

See Appendix \ref{app:inthamt} for the outline of the derivation of Eq. (\ref{fermifermiintHt}).

\subsection{Master equation}
\label{subsec:mastereqn}

As stated earlier, the interaction of the fermions trapped in the lattice with a bath of fermionic particles causes them to either get trapped by or expelled from the lattice system over time. Our task is now to determine the time evolution equation for the fermions in the lattice.

The assumptions made in deriving the equation are: (1) the system has a negligible effect on the bath and (2) the state of the system at a particular time step depends only on its state in the previous time step. These assumptions constitute the Born-Markov approximation. The time evolution of systems that follow the Born-Markov approximation is described by the equation \cite{bp}
\begin{equation} \label{bornmarkovdef}
	\frac{d}{dt} \rho_S (t) = - \int_{0}^{\infty} ds Tr_B [H_I (t-s). [H_I(s), \rho_S (t) \otimes \rho_B]].
\end{equation}

Substituting the interaction Hamiltonian in Eq. (\ref{intHfermi}) and using the anticommutator relations of the fermionic operators \cite{mahan}
\begin{eqnarray}
\{ \hat{c}_{ \vec{l}, \sigma }, \hat{c}_{ \vec{l'}, \sigma' } \} &=& \{ \hat{c}_{ \vec{l}, \sigma }^{\dagger}, \hat{c}_{ \vec{l'}, \sigma' }^{\dagger} \} = 0 \label{anticomm1} \\
\{ \hat{c}_{ \vec{l}, \sigma }, \hat{c}_{ \vec{l'}, \sigma' }^{\dagger} \} &=& \delta_{ \vec{l}, \vec{l'} } \delta_{ \sigma, \sigma' } \label{anticomm2}
\end{eqnarray}
together with the rotating wave approximation, we then find that the master equation of the system in this study is
\begin{widetext}
	\begin{align}
	\frac{d}{dt} \rho_S (t) = \frac{2 \pi^2 a_S^2 \rho_C}{i V \delta_R \mu_R^2} \sum_{\vec{l}, \vec{l}'} & \sum_{\vec{m}} A_{\vec{m}, \vec{l}}^2 \left( u_{\vec{m}} + v_{\vec{m}} \right)^2 \nonumber \\
	& \times \{ \left( [\hat{c}_{\vec{l'}, \downarrow}^\dagger \hat{c}_{\vec{l'} + \vec{r'}, \downarrow}, [\hat{c}_{\vec{l}, \uparrow}^\dagger \hat{c}_{\vec{l} + \vec{r}, \uparrow}, \rho_S (t)] ] - [\hat{c}_{\vec{l'}, \uparrow}^\dagger \hat{c}_{\vec{l'} + \vec{r'}, \uparrow}, [\hat{c}_{\vec{l}, \downarrow}^\dagger \hat{c}_{\vec{l} + \vec{r}, \downarrow}, \rho_S (t)] ] + h.c. \right) \nonumber \\
	& + \left( [\hat{c}_{\vec{l'}, \downarrow}^\dagger \hat{c}_{\vec{l'} + \vec{r'}, \downarrow}, [\hat{c}_{\vec{l} + \vec{r}, \downarrow}^\dagger \hat{c}_{\vec{l}, \downarrow}, \rho_S (t)] ] - [\hat{c}_{\vec{l'}, \uparrow}^\dagger \hat{c}_{\vec{l'} + \vec{r'}, \uparrow}, [\hat{c}_{\vec{l} + \vec{r}, \uparrow}^\dagger \hat{c}_{\vec{l}, \uparrow}, \rho_S (t)] ] - h.c. \right) \}.
	\label{fermifermimasteqn}
	\end{align}
\end{widetext}

See Appendix \ref{app:mastereqn} for the detailed outline of the derivation master equation.

\section{Information Transport}
\label{sec:occprobfid}

\subsection{System and initial conditions}
\label{subsec:systeminitcond}

With the dynamics of particles in the lattice described by Eq. (\ref{fermifermimasteqn}), the system can be simulated numerically.

There are two types of lattices considered for the Fermi-Hubbard model. First is an open lattice where the first and the third sites do not interact directly (see Figure \ref{fig:schematicopen}). Second is a closed lattice where there is nearest neighbor interaction between the sites, including the first and third (see Figure \ref{fig:schematicclosed}).

\subsection{Fidelity}
\label{subsec:fidelity}

\begin{figure*}[!tbp]
	\centering
	\subfigure[Open; empty lattice.]{\includegraphics[scale=0.35]{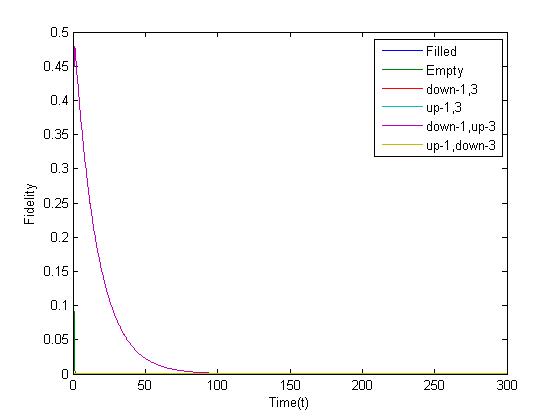}\label{fig:3openemptyfidelity}}
	\quad
	\subfigure[Closed; empty lattice.]{\includegraphics[scale=0.35]{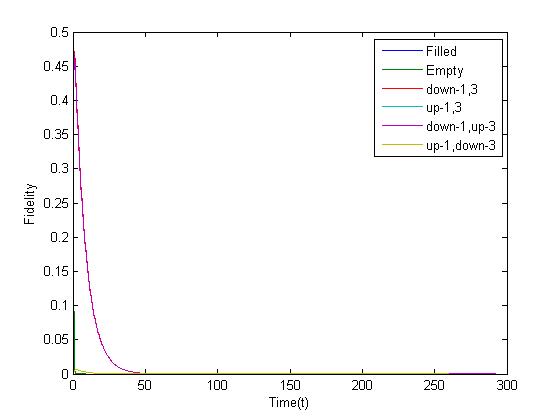}\label{fig:3closedemptyfidelity}}
	\subfigure[Open; maximally entangled.]{\includegraphics[scale=0.35]{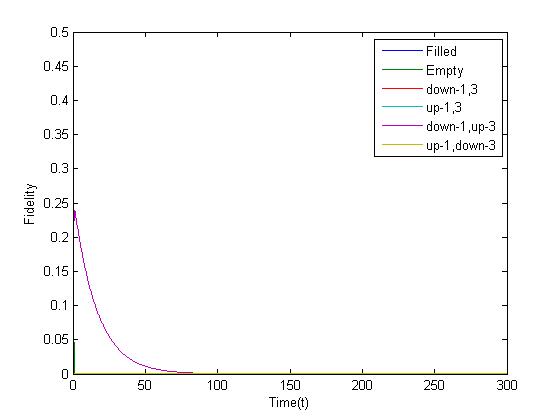}\label{fig:3openmaxentfidelity}}
	\quad
	\subfigure[Closed; maximally entangled.]{\includegraphics[scale=0.35]{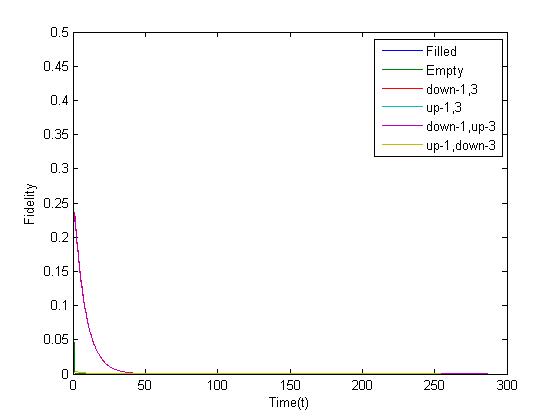}\label{fig:3closedmaxentfidelity}}
	\subfigure[Open; 30\% spin-down in first site, 70\% spin-down in third site.]{\includegraphics[scale=0.35]{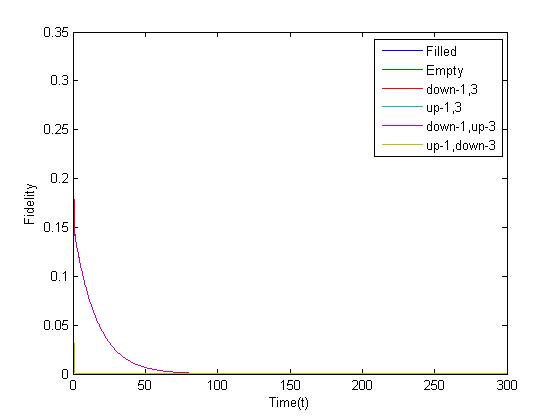}\label{fig:3open3070fidelity}}
	\quad
	\subfigure[Closed; 30\% spin-down in first site, 70\% spin-down in third site.]{\includegraphics[scale=0.35]{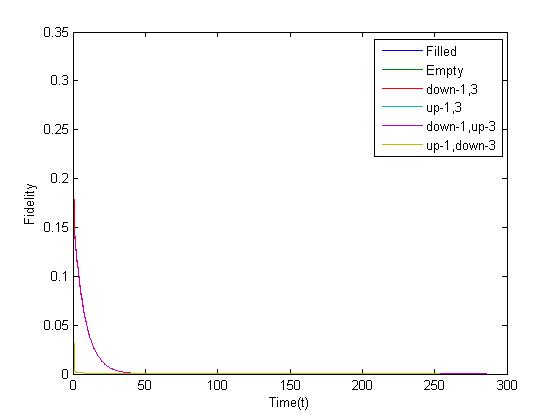}\label{fig:3closed3070fidelity}}
	\subfigure[Open; 70\% spin-down in first site, 30\% spin-down in third site.]{\includegraphics[scale=0.35]{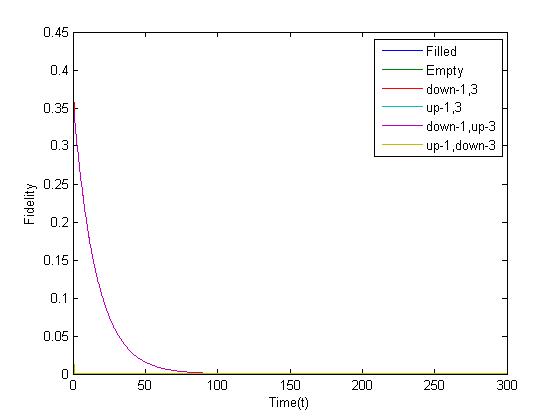}\label{fig:3open7030fidelity}}
	\quad
	\subfigure[Closed; 70\% spin-down in first site, 30\% spin-down in third site.]{\includegraphics[scale=0.35]{3closed3070fidelity}\label{fig:3closed7030fidelity}}
	\caption{Fidelity with respect to different final states.}
	\label{fig:fidelitycomp1}	
\end{figure*}

To describe the behavior of the gas in the lattice, we can observe its probability of occupying a state $\rho_f$,
\begin{equation}
F (t) = Tr(\rho_f \rho_S (t)),
\end{equation}
known as the fidelity. From the state of the system, we can know if information is being transferred from one lattice site through the particles and their specific spins.

To observe the fidelity (and the occupation probability and the entanglement in the system), we denote four different initial states of the system:
\begin{enumerate}
	\item{Pure state in which the lattice has no particles:
		\begin{equation*}
		\ket{0_{\uparrow}, 0_{\downarrow}}_1 \ket{0_{\uparrow}, 0_{\downarrow}}_2 \ket{0_{\uparrow}, 0_{\downarrow}}_3 \bra{0_{1\uparrow}, 0_{1\downarrow}}_1 \bra{0_{\uparrow}, 0_{\downarrow}}_2 \bra{0_{\uparrow}, 0_{\downarrow}}_3
		\end{equation*}}
	\item{Maximally entangled state such that the lattice is half-empty and half-full:
		\begin{eqnarray}
		\frac{1}{2} \ket{0_{\uparrow}, 0_{\downarrow}}_1 \ket{0_{\uparrow}, 0_{\downarrow}}_2 \ket{0_{\uparrow}, 0_{\downarrow}}_3 \bra{0_{\uparrow}, 0_{\downarrow}}_1 \bra{0_{\uparrow}, 0_{\downarrow}}_2 \bra{0_{\uparrow}, 0_{\downarrow}}_3 && \nonumber \\
		+ \frac{1}{2} \ket{1_{\uparrow}, 1_{\downarrow}}_1 \ket{1_{\uparrow}, 1_{\downarrow}}_2 \ket{1_{\uparrow}, 1_{\downarrow}}_3 \bra{1_{\uparrow}, 1_{\downarrow}}_1 \bra{1_{\uparrow}, 1_{\downarrow}}_2 \bra{1_{\uparrow}, 1_{\downarrow}}_3 && \nonumber
		\end{eqnarray}}
	\item{70\% spin-down in the first site and 30\% spin-down in the third site:
		\begin{eqnarray}
		\frac{7}{10} \ket{0_{\uparrow}, 1_{\downarrow}}_1 \ket{0_{\uparrow}, 0_{2, \downarrow}}_2 \ket{0_{\uparrow}, 0_{\downarrow}}_3 \bra{0_{\uparrow}, 1_{\downarrow}}_1 \bra{0_{\uparrow}, 0_{\downarrow}}_2 \bra{0_{\uparrow}, 0_{\downarrow}}_3 && \nonumber \\
		+ \frac{3}{10} \ket{0_{\uparrow}, 0_{\downarrow}}_1 \ket{0_{\uparrow}, 0_{\downarrow}}_2 \ket{0_{\uparrow}, 1_{\downarrow}}_3 \bra{0_{\uparrow}, 0_{\downarrow}}_1 \bra{0_{\uparrow}, 0_{\downarrow}}_2 \bra{0_{\uparrow}, 1_{\downarrow}}_3 && \nonumber
		\end{eqnarray}}
	\item{30\% spin-down in the first site and 70\% spin-down in the third site:
		\begin{eqnarray}
		\frac{3}{10} \ket{0_{\uparrow}, 1_{\downarrow}}_1 \ket{0_{\uparrow}, 0_{2, \downarrow}}_2 \ket{0_{\uparrow}, 0_{\downarrow}}_3 \bra{0_{\uparrow}, 1_{\downarrow}}_1 \bra{0_{\uparrow}, 0_{\downarrow}}_2 \bra{0_{\uparrow}, 0_{\downarrow}}_3 && \nonumber \\
		+ \frac{7}{10} \ket{0_{\uparrow}, 0_{\downarrow}}_1 \ket{0_{\uparrow}, 0_{\downarrow}}_2 \ket{0_{\uparrow}, 1_{\downarrow}}_3 \bra{0_{\uparrow}, 0_{\downarrow}}_1 \bra{0_{\uparrow}, 0_{\downarrow}}_2 \bra{0_{\uparrow}, 1_{\downarrow}}_3 && \nonumber
		\end{eqnarray}}
\end{enumerate}

Figure \ref{fig:fidelitycomp1} summarizes the results of computations of the fidelity for cases of different initial states. The states $\rho_f$ with respect to which the fidelity is computed are the following:
\begin{enumerate}
	\item{the lattice is empty:
		\begin{equation*}
		\ket{0_{\uparrow}, 0_{\downarrow}}_1 \ket{0_{\uparrow}, 0_{\downarrow}}_2 \ket{0_{\uparrow}, 0_{\downarrow}}_3 \bra{0_{1\uparrow}, 0_{1\downarrow}}_1 \bra{0_{\uparrow}, 0_{\downarrow}}_2 \bra{0_{\uparrow}, 0_{\downarrow}}_3
		\end{equation*}}
	\item{the lattice is full:
		\begin{equation*}
		\ket{1_{\uparrow}, 1_{\downarrow}}_1 \ket{1_{\uparrow}, 1_{\downarrow}}_2 \ket{1_{\uparrow}, 1_{\downarrow}}_3 \bra{1_{\uparrow}, 1_{\downarrow}}_1 \bra{1_{\uparrow}, 1_{\downarrow}}_2 \bra{1_{\uparrow}, 1_{\downarrow}}_3
		\end{equation*}}
	\item{50\% probability that a spin-down particle is found in the first site and 50\% probability that a spin-down particle is found in the third site:
		\begin{eqnarray}
		\frac{1}{2} \ket{0_{\uparrow}, 1_{\downarrow}}_1 \ket{0_{\uparrow}, 0_{\downarrow}}_2 \ket{0_{\uparrow}, 0_{\downarrow}}_3 \bra{0_{\uparrow}, 1_{\downarrow}}_1 \bra{0_{\uparrow}, 0_{\downarrow}}_2 \bra{0_{\uparrow}, 0_{\downarrow}}_3 && \nonumber \\
		+ \frac{1}{2} \ket{0_{\uparrow}, 0_{\downarrow}}_1 \ket{0_{\uparrow}, 0_{\downarrow}}_2 \ket{0_{\uparrow}, 1_{\downarrow}}_3 \bra{0_{\uparrow}, 0_{\downarrow}}_1 \bra{0_{\uparrow}, 0_{\downarrow}}_2 \bra{0_{\uparrow}, 1_{\downarrow}}_3 && \nonumber
		\end{eqnarray}}
	\item{50\% probability that a spin-up particle is found in the first site and 50\% probability that a spin-up particle is found in the third site:
		\begin{eqnarray}
		\frac{1}{2} \ket{1_{\uparrow}, 0_{\downarrow}}_1 \ket{0_{\uparrow}, 0_{\downarrow}}_2 \ket{0_{\uparrow}, 0_{\downarrow}}_3 \bra{1_{\uparrow}, 0_{\downarrow}}_1 \bra{0_{\uparrow}, 0_{\downarrow}}_2 \bra{0_{\uparrow}, 0_{\downarrow}}_3 && \nonumber \\
		+ \frac{1}{2} \ket{0_{\uparrow}, 0_{\downarrow}}_1 \ket{0_{\uparrow}, 0_{\downarrow}}_2 \ket{1_{\uparrow}, 0_{\downarrow}}_3 \bra{0_{\uparrow}, 0_{\downarrow}}_1 \bra{0_{\uparrow}, 0_{\downarrow}}_2 \bra{1_{\uparrow}, 0_{\downarrow}}_3 && \nonumber
		\end{eqnarray}}
	\item{50\% probability that a spin-down particle is found in the first site and 50\% probability that a spin-up particle is found in the third site:
		\begin{eqnarray}
		\frac{1}{2} \ket{0_{\uparrow}, 1_{\downarrow}}_1 \ket{0_{\uparrow}, 0_{2, \downarrow}}_2 \ket{0_{\uparrow}, 0_{\downarrow}}_3 \bra{0_{\uparrow}, 1_{\downarrow}}_1 \bra{0_{\uparrow}, 0_{\downarrow}}_2 \bra{0_{\uparrow}, 0_{\downarrow}}_3 && \nonumber \\
		+ \frac{1}{2} \ket{0_{\uparrow}, 0_{\downarrow}}_1 \ket{0_{\uparrow}, 0_{\downarrow}}_2 \ket{0_{\uparrow}, 1_{\downarrow}}_3 \bra{0_{\uparrow}, 0_{\downarrow}}_1 \bra{0_{\uparrow}, 0_{\downarrow}}_2 \bra{0_{\uparrow}, 1_{\downarrow}}_3 && \nonumber
		\end{eqnarray}}
	\item{50\% probability that a spin-up particle is found in the first site and 50\% probability that a spin-down particle is found in the third site:
		\begin{eqnarray}
		\frac{1}{2} \ket{1_{\uparrow}, 0_{\downarrow}}_1 \ket{0_{\uparrow}, 0_{2, \downarrow}}_2 \ket{0_{\uparrow}, 0_{\downarrow}}_3 \bra{1_{\uparrow}, 0_{\downarrow}}_1 \bra{0_{\uparrow}, 0_{\downarrow}}_2 \bra{0_{\uparrow}, 0_{\downarrow}}_3 && \nonumber \\
		+ \frac{1}{2} \ket{0_{\uparrow}, 0_{\downarrow}}_1 \ket{0_{\uparrow}, 0_{\downarrow}}_2 \ket{1_{\uparrow}, 0_{\downarrow}}_3 \bra{0_{\uparrow}, 0_{\downarrow}}_1 \bra{0_{\uparrow}, 0_{\downarrow}}_2 \bra{1_{\uparrow}, 0_{\downarrow}}_3 && \nonumber
		\end{eqnarray}}
\end{enumerate}
The choice of these states are based on the following: the first two $\rho_f$s tell if the whole lattice would be filled with information while the latter four $\rho_f$s (particularly the application of the $6^{th}$ $\rho_f$ on the $3^{rd}$ and $4^{th}$ initial state cases indicated earlier) would indicate if a spin-exchange happened or not. 

We can then see in Figures \ref{fig:3openemptyfidelity} and \ref{fig:3closedemptyfidelity} that an empty lattice does not remain empty through time for there is a nonzero probability that at least one spin-up and one spin-down particle occupies either the first or the third lattice site at some time $t \neq 0$.

On the other hand, when the system is initially in the maximally entangled state where there is 50\% probability of finding the lattice empty and 50\% probability of finding the lattice full of particles as seen in Figures \ref{fig:3openmaxentfidelity} and \ref{fig:3closedmaxentfidelity}, the 50\% chance of finding the system full is constant though time. However, the probability that the lattice is found empty becomes zero at $t > 0$ and that the other states are occupied.

Lastly, we consider the cases wherein there is only one particle in the lattice and there is a nonzero probability, (either 30\% or 70\%) that it is in the first site or in the third site. The results in Figures \ref{fig:3open3070fidelity} to \ref{fig:3closed7030fidelity} show the fidelity such that there is 50\% probability of finding a spin-down particle in the first site and 50\% probability of finding a spin-up particle in the third site.

Through the numerical results, we can draw the conclusion that fermionic particles from the bath enter into and exit from the lattice through time. Spin exchange can also happen such that the spin of the particle on either the first or third lattice flips from down to up at time $t > 0$.

\subsection{Occupation probability}
\label{subsec:occprob}

\begin{figure*}[!tbp]
	\centering
	\subfigure[Open; empty lattice.]{\includegraphics[scale=0.35]{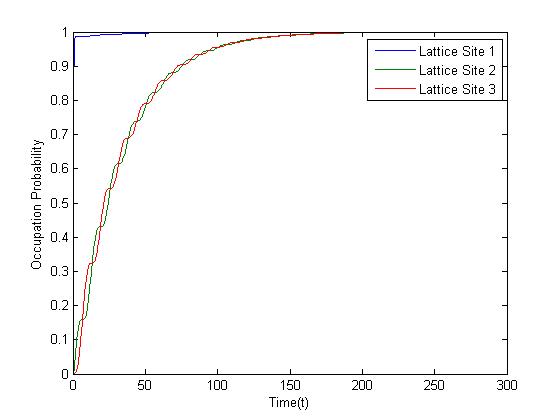}\label{fig:3openemptyoccprob}}
	\quad
	\subfigure[Closed; empty lattice.]{\includegraphics[scale=0.35]{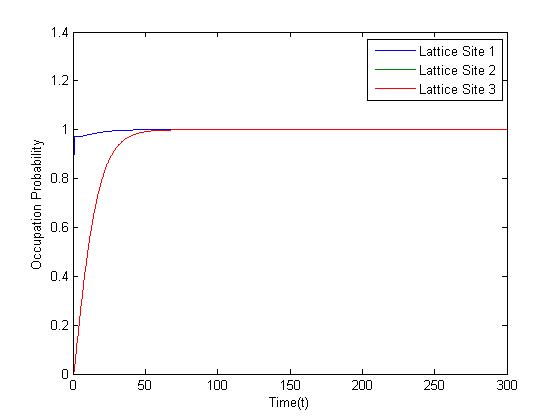}\label{fig:3closedemptyoccprob}}
	\subfigure[Open; maximally entangled.]{\includegraphics[scale=0.35]{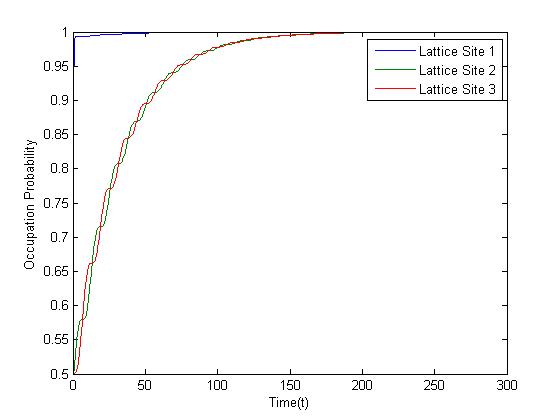}\label{fig:3openmaxentoccprob}}
	\quad
	\subfigure[Closed; maximally entangled.]{\includegraphics[scale=0.35]{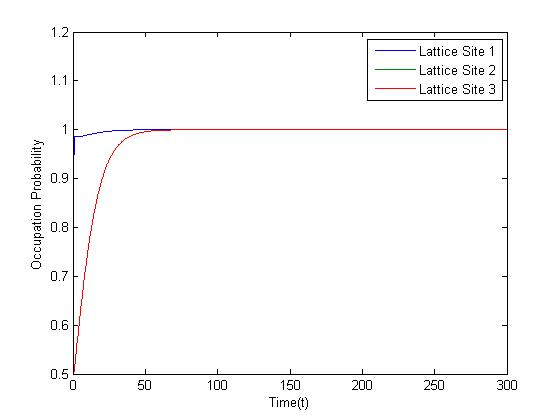}\label{fig:3closedmaxentoccprob}}
	\subfigure[Open; 30\% spin-down in first site, 70\% spin-down in third site.]{\includegraphics[scale=0.35]{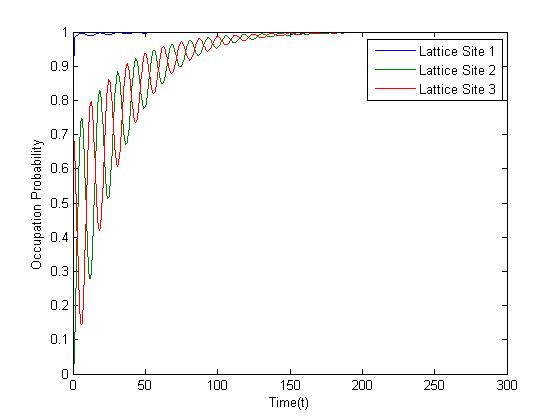}\label{fig:3open3070occprob}}
	\quad
	\subfigure[Closed; 30\% spin-down in first site, 70\% spin-down in third site.]{\includegraphics[scale=0.35]{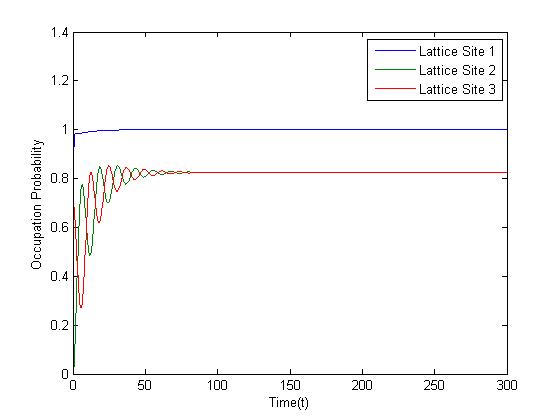}\label{fig:3closed3070occprob}}
	\subfigure[Open; 70\% spin-down in first site, 30\% spin-down in third site.]{\includegraphics[scale=0.35]{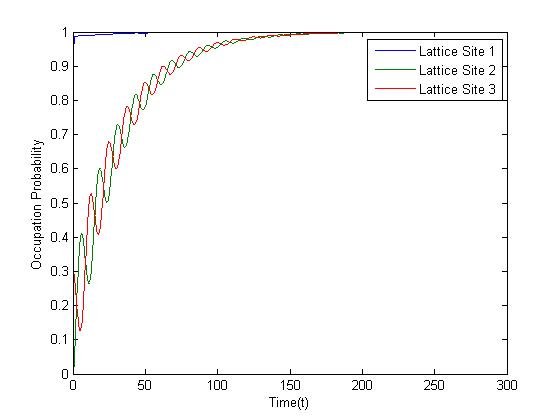}\label{fig:3open7030occprob}}\quad
	\subfigure[Closed; 70\% spin-down in first site, 30\% spin-down in third site.]{\includegraphics[scale=0.35]{3closed3070occprob}\label{fig:3closed7030occprob}}
	\caption{Occupation probability of each lattice site.}
	\label{fig:occprob}	
\end{figure*}

Another quantity that describes the dynamics of the particles in the lattice through time is the occupation probability. This is the probability in which at least one particle, regardless of spin, occupies the lattice site. For the $n^{th}$ site, the equation for the occupation probability is
\begin{equation}
	P_{occ, n} (t) = Tr(\rho_S (t) \rho_{occ,n})
	\label{occprob}
\end{equation}
where $\rho_{occ,n}$ a $64 \times 64$ diagonal matrix where any entry pertaining to non-occupation of the $n^{th}$ lattice site is zero and the rest is equal to one. For each lattice site, it can be stated as follows:
\begin{widetext}
	\label{occprobforms}
	\begin{align}
	\rho_{occ, 1} = \sum_{n_{2, \uparrow}, n_{2, \downarrow}} \sum_{n_{3, \uparrow}, n_{3, \downarrow}} & ( \ket{1_{\uparrow}, 1_{\downarrow}}_1 \bra{1_{\uparrow}, 1_{\downarrow}}_1 + \ket{1_{\uparrow}, 0_{\downarrow}}_1 \bra{1_{\uparrow}, 0_{\downarrow}}_1 + \ket{0_{\uparrow}, 1_{\downarrow}}_1 \bra{0_{\uparrow}, 1_{\downarrow}}_1 ) \nonumber \\
	& \otimes \ket{n_{2, \uparrow}, n_{2, \downarrow}}_2 \bra{n_{2, \uparrow}, n_{2, \downarrow}}_2 \otimes \ket{n_{3, \uparrow}, n_{3, \downarrow}}_3 \bra{n_{3, \uparrow}, n_{3, \downarrow}}_3 \\
	\rho_{occ, 2} = \sum_{n_{1, \uparrow}, n_{1, \downarrow}} \sum_{n_{3, \uparrow}, n_{3, \downarrow}} & \ket{n_{1, \uparrow}, n_{1, \downarrow}}_1 \bra{n_{1, \uparrow}, n_{1, \downarrow}}_1 \otimes ( \ket{1_{\uparrow}, 1_{\downarrow}}_2 \bra{1_{\uparrow}, 1_{\downarrow}}_2 + \ket{1_{\uparrow}, 0_{\downarrow}}_2 \bra{1_{\uparrow}, 0_{\downarrow}}_2 \nonumber \\
	& + \ket{0_{\uparrow}, 1_{\downarrow}}_2 \bra{0_{\uparrow}, 1_{\downarrow}}_2 ) \otimes \ket{n_{3, \uparrow}, n_{3, \downarrow}}_3 \bra{n_{3, \uparrow}, n_{3, \downarrow}}_3 \\
	\rho_{occ, 3} = \sum_{n_{1, \uparrow}, n_{1, \downarrow}} \sum_{n_{2, \uparrow}, n_{2, \downarrow}} & \ket{n_{1, \uparrow}, n_{1, \downarrow}}_1 \bra{n_{1, \uparrow}, n_{1, \downarrow}}_1 \otimes \ket{n_{2, \uparrow}, n_{2, \downarrow}}_2 \bra{n_{2, \uparrow}, n_{2, \downarrow}}_2 \nonumber \\
	& \otimes ( \ket{1_{\uparrow}, 1_{\downarrow}}_2 \bra{1_{\uparrow}, 1_{\downarrow}}_2 + \ket{1_{\uparrow}, 0_{\downarrow}}_2 \bra{1_{\uparrow}, 0_{\downarrow}}_2 + \ket{0_{\uparrow}, 1_{\downarrow}}_2 \bra{0_{\uparrow}, 1_{\downarrow}}_2 ),
	\end{align}
\end{widetext}
where $n_{\vec{j}, \sigma} \in \{ 0, 1 \}$.

The sum of all the occupation probabilities is not normalized because the individual sites, not the lattice as a whole, are observed.

As seen in Figure \ref{fig:occprob}, for all cases, there is a nonzero probability that at least one particle, regardless of spin, would occupy each lattice site at $t > 0$ even if the lattice is initially empty. This confirms the statement in Section \ref{subsec:fidelity} that the lattice does not remain empty through time as observed using fidelity.

For the first two initial conditions as seen in Figures \ref{fig:3openemptyoccprob} to \ref{fig:3closedmaxentoccprob}, the occupation probability of each lattice site increases through time until it approaches 1 as $t \rightarrow \infty$. The occupation probability of the first site, however, is the one that approaches the value of unity fastest among the three sites.

On the other hand, as seen through Figures \ref{fig:3open3070occprob} to \ref{fig:3closed7030occprob}, the occupation probabilities of all sites stabilize at unity for open lattices. However, for closed lattices, only the occupation probability of the first site approaches 1 as $t \rightarrow \infty$ while the occupation probability of the other sites approach $0.83$ as $t \rightarrow \infty$. This means that there is a chance that no particle is present in the second and third site for long times.

\subsection{Entanglement}
\label{subsec:entanglement}

\begin{figure*}[!tbp]
	\centering
	\subfigure[Open; empty lattice.]{\includegraphics[scale=0.35]{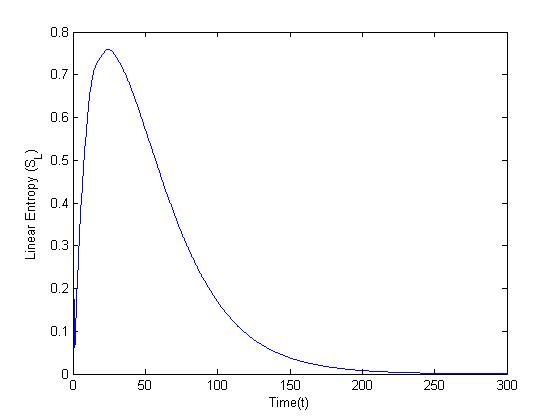}\label{fig:3openemptyent}}
	\quad
	\subfigure[Closed; empty lattice.]{\includegraphics[scale=0.35]{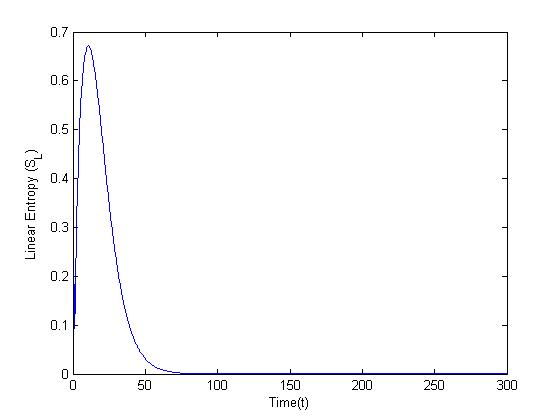}\label{fig:3closedemptyent}}
	\subfigure[Open; maximally entangled.]{\includegraphics[scale=0.35]{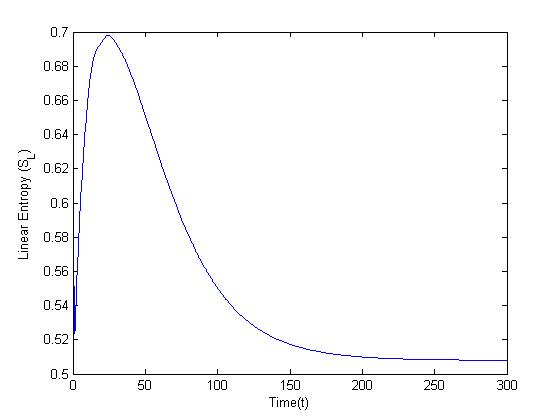}\label{fig:3openmaxentent}}
	\quad
	\subfigure[Closed; maximally entangled.]{\includegraphics[scale=0.35]{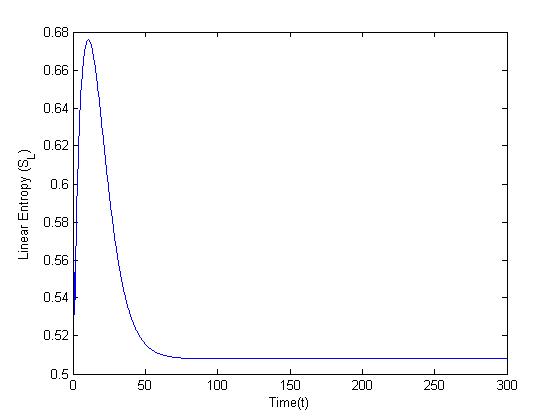}\label{fig:3closedmaxentent}}
	\subfigure[Open; 30\% spin-down in first site, 70\% spin-down in third site.]{\includegraphics[scale=0.35]{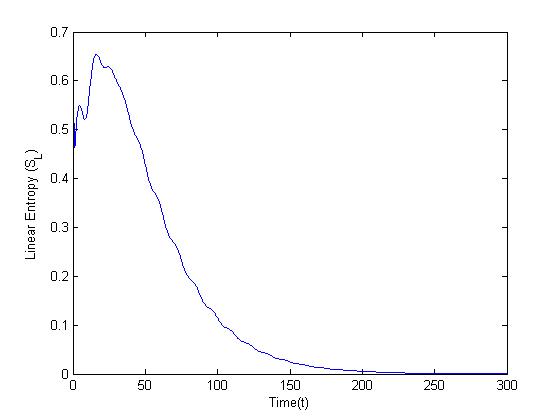}\label{fig:3open3070ent}}
	\quad
	\subfigure[Closed; 30\% spin-down in first site, 70\% spin-down in third site.]{\includegraphics[scale=0.35]{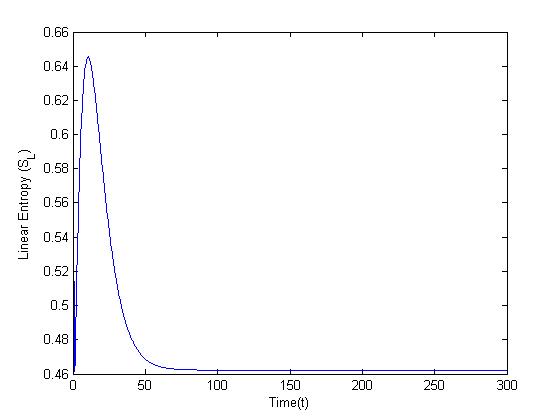}\label{fig:3closed3070ent}}
	\subfigure[Open; 70\% spin-down in first site, 30\% spin-down in third site.]{\includegraphics[scale=0.35]{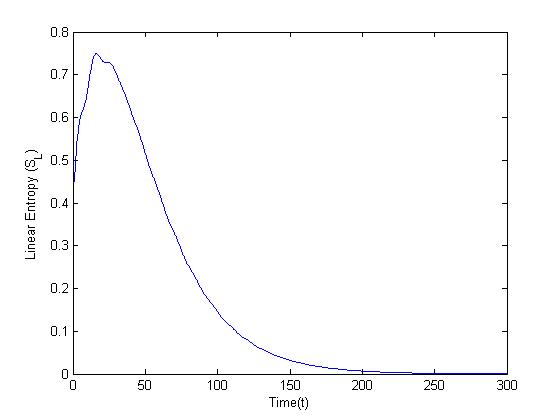}\label{fig:3open7030ent}}\quad
	\subfigure[Closed; 70\% spin-down in first site, 30\% spin-down in third site.]{\includegraphics[scale=0.35]{3closed3070ent}\label{fig:3closed7030ent}}
	\caption{Linear entropy with respect to different initial states.}
	\label{fig:entropy}	
\end{figure*}

Lastly, the transport of information through the lattice can be described by measuring the entanglement of the optical lattice with entropy.

In this study, the measure of entanglement is the linear entropy,
\begin{equation} \label{ent}
S_L = \frac{d}{d-1}(1 - Tr(\rho_S (t)^2)),
\end{equation}
wherein $d$ is the dimension of the system, i.e. the number of the lattice sites. (In this study, $d=64$.) Linear entropy was chosen as the entanglement measure in this study as it is measure that can be applicable to any system of any dimension (as measures such as concurrence fail for higher dimensions).

It is then seen in Figure \ref{fig:entropy} that in general, entanglement as measured by the linear entropy of the time-evolved state peaks at some time $t > 0$ then decreases, stabilizing at a certain value as $t \rightarrow \infty$.

For open lattices, we can consider as an illustrative case that shown in Figure \ref{fig:3openmaxentent}, where the lattice system is considered as maximally entangled at $t=0$ as there is a 50\% probability of finding the system full of fermionic particles and 50\% probability of finding it empty. Comparing this to the other cases for the open lattice, this shows that if the initial state of the lattice has a nonzero probability of having the lattice full of particles, nonzero linear entropy can be achieved as $t \rightarrow \infty$. This means that the initial probability of having the lattice full can partially preserve the entanglement through time.

On the other hand, for closed lattices, the linear entropy $S_L \neq 0$ as $t \rightarrow \infty$ in general. However, the linear entropy stabilizes at a maximum value of $0.51$ as $t \rightarrow \infty$ compared to the other cases with respect to the initial condition. Therefore, in general, closing the lattice and starting the lattice in a state such that there is a nonzero initial probability of finding the lattice full of particles helps increase the long-time value of entanglement in the system.

\section{Summary}
\label{sec:conclusion}

In this article, we examine the dissipative dynamics of a Fermi gas in a three-site optical lattice exposed to a fermionic environment as illustrated in Figure \ref{fig:schematic}. Since fermion-fermion scattering takes place due to the interaction of the lattice and environment, intrinsic properties of the particle, i.e. spin, are regarded as information passed through the lattice.

Using the fidelity and the occupation probability, we find that even if at $t = 0$, the lattice is empty, particles from the environment will enter into it. Furthermore, we observe through the fidelity that spin exchange happens in the system such that a spin-up particle can turn into a spin-down particle upon the exposure of the system to the fermionic bath. Lastly, closing the lattice and starting it in a state such that there is nonzero probability of finding the lattice full helps preserve a higher value of entanglement.

The results imply that the three-site optical lattice system can be used to realize quantum technologies, such as quantum wires or a series of quantum computers by means of quantum transport of information through lattice sites, which may be parts of a quantum wire or nodes in a network of quantum computers. Furthermore, the system can be used for quantum encryption via spin exchange within a quantum or classical network.

This work is limited to a three-site optical lattice due to limitations in available computing resources. However, this work can be extended to a lattice with four or more sites, which we intend to do in future work.

\section*{Acknowledgement}

This research is funded by a grant from the National Research Council of the Philippines (NRCP) as NRCP Project P-022. Roland Caballar would like to acknolwedge M. A. A. Estrella for valuable technical and conceptual assistance during the research. Vladimir Villegas would like to acknowledge the financial support from the Department of Science and Technology (DOST) via the Advanced Science and Technology Human Resource Development (ASTHRDP), J. L. Duanan for valuable technical assistance and B. Villegas, E. Galapon, R. J. Abuel, G.-G. Flores, and D. M. Antazo for valuable conceptual assistance during the research. Finally, V. Villegas and R. Caballar would like to acknowledge T. Lindberg, A. Bj{\"o}rler, J. Bj{\"o}rler, M. Larsson, A. Erlandsson, B. Hannigan, A. Netrebko, K. Te Kanawa, D. Damrau, C. Bartoli, T. Bangalter, G-M. de Homem-Christo, E. L. B. Gil, and H. E. Soberano for conceptual discussions.

\appendix

\section{Properties of the Wannier functions}
\label{app:wannier}

The Wannier functions $w_0 (\vec{x})$ have the following properties.
\begin{enumerate}
	\item{An alternative notation is as follows:
		\begin{equation}
		w_0 (\vec{r} - \vec{R}) = w_{0 \vec{R}} (\vec{r}).
		\end{equation}}
	\item{It then follows that translation of the Wannier functions would be
		\begin{equation}
		w_{0 \vec{R}} (\vec{r}) = w_{0 \vec{R} + \vec{R'}} (\vec{r} + \vec{R'}).
		\end{equation}}
	\item{The Bloch functions $\phi_{\vec{k}} (\vec{r})$ can be expressed in terms of Wannier functions through the Fourier transform
		\begin{equation}
		\phi_{\vec{k}} (\vec{r}) = \frac{1}{N} \sum_{\vec{R}} e^{i \vec{k} \cdot \vec{R}} w_0 (\vec{r} - \vec{R})
		\end{equation}
	wherein $N$ is a normalization constant.}
	\item{Using the expression of Wannier functions in terms of Bloch functions, we know that the Wannier functions are orthonormal, i.e.
		\begin{equation}
		\int_{crystal} d^3 \vec{r} w_0 (\vec{r} - \vec{R}) w_0 (\vec{r} - \vec{R'}) = \delta_{\vec{R},\vec{R'}}.
		\end{equation}}
\end{enumerate}

\section{Derivation of the interaction Hamiltonian}
\label{app:intham}

The interaction Hamiltonian is expressed in terms of the lattice and bath field operators in Eq. (\ref{inthamgen}). However, we can write the interaction Hamiltonian in terms of the lattice and bath annihilation and creation operators in order to express the effects of the particles on the bath onto the lattice. We substitute the field operators of the optical lattice in Eq. (\ref{latoptr}) and the field operators of the fermionic bath in (\ref{bathfieldop}), we get the following.
\begin{widetext}
	\begin{align}
	H_I = \frac{2 \pi a \rho_C}{\mu_R} \int & d^3 \vec{x} \left( \sum_{\vec{l}, \sigma} \hat{c}_{\vec{l}, \sigma}^\dagger w_0 (\vec{x} - \vec{x}_{\vec{l}}) \right) \left( \sum_{\vec{l'}, \sigma'} \hat{c}_{\vec{l'}, \sigma'} w_0 (\vec{x} - \vec{x}_{\vec{l'}}) \right) \cdot \left( \sqrt{\rho_C} + \frac{1}{V} \sum_{\vec{m}} \sum_{\alpha} (u_{\vec{m}} \hat{h}_{\vec{m}, \sigma}^\dagger e^{-i \vec{m} \cdot \vec{x}} + v_{\vec{m}} \hat{h}_{\vec{m}, \sigma} e^{i \vec{m} \cdot \vec{x}}) \right) \nonumber \\
	& \left( \sqrt{\rho_C} + \frac{1}{V} \sum_{\vec{m'}} \sum_{\alpha'} (u_{\vec{m'}} \hat{h}_{\vec{m'}, \alpha'} e^{i \vec{m'} \cdot \vec{x}} + v_{\vec{m'}} \hat{h}_{\vec{m'}, \alpha'}^{\dagger} e^{-i \vec{m'} \cdot \vec{x}}) \right)
	\end{align}
\end{widetext}

Multiplying the lattice field operators with the first terms of the bath field operators would just result to orthonormality conditions. However, note that we can only apply orthonormality on the position but not on the spin.

On the other hand, for terms with bath creation and annihilation operators, we use the properties of the Wannier functions to evaluate the integrals and drop terms of the order $\hat{h}_{\vec{m}, \alpha}^{(\dagger)} \hat{h}_{\vec{m'}, \alpha'}^{(\dagger)}$ (since we restrict ourselves to two-body interaction).

The overlap integral is expressed as \cite{caballar}
\begin{equation}
A_{\vec{m},\vec{l}} = e^{i \vec{l} \cdot \vec{x}_{\vec{l}}} e^{-i (\vec{l} + \vec{r}) \cdot \vec{x}_{\vec{l} + \vec{r}}} \int d^3 \vec{x} e^{i \vec{m} \cdot \vec{x}} \phi^{\ast}_{\vec{l}} ( \vec{x} ) \phi_{\vec{l} + \vec{r}} ( \vec{x} ).
\end{equation}

Since the interaction Hamiltonian must be Hermitian, then we know that $A_{\vec{m},\vec{l}}$ is real. Therefore, we can also impose that $A_{\vec{m},\vec{l}} = A_{-\vec{m},\vec{l}}$.

Therefore, we have Eq. (\ref{intHfermi}).

\section{Derivation of the time-evolved interaction Hamiltonian}
\label{app:inthamt}

After the derivation of the interaction Hamiltonian, the next step is to evolve it through time. It can be done using the Baker-Campbell-Hausdorff formula.
\begin{eqnarray}
H_I (t) &=& e^{i H_0 t} H_I e^{-i H_0 t}; (\hbar \equiv 1) \\
&=& 1 + i \left[ H_0, H_I \right] - \frac{1}{2} \left[ H_0, \left[ H_0, H_I \right] \right] + ...
\label{bch}
\end{eqnarray}
$H_0$ is the free Hamiltonian given in Eq. (\ref{freeHop}) and $H_I$ is the interaction Hamiltonian given in Eq. (\ref{intHfermi}).

The Pauli exclusion principle states that only one particle can occupy an energy level. In our case, since we are particular with the spin, two fermions - one spin-up and one spin-down - can occupy the same lattice site at a time.

Therefore, in evaluating the nested commutators in Eq. (\ref{bch}), we make use of the anticommutator relations given in Eq. (\ref{anticomm1}) and Eq. (\ref{anticomm2}) \cite{mahan}.

On the other hand, we are just concerned with two-body interactions. Therefore, we can neglect terms with three or more fermionic operators.

Therefore, the time-evolved interaction Hamiltonian for the fermion-fermion scattering case is Eq. (\ref{fermifermiintHt}).

\section{Derivation of the master equation}
\label{app:mastereqn}

In this study, we apply the Born-Markov approximation, meaning that the action of the optical lattice on the fermionic bath is negligible such that they have minimal coupling and the state at time $t$ depends only on the immediate previous time step. Therefore, the dynamics of the system is guided by the Born-Markov master equation given in Eq. (\ref{bornmarkovdef}) \cite{bp}.

Substituting Eq. (\ref{fermifermiintHt}) into Eq. (\ref{bornmarkovdef}), we obtain a series of two nested commutators. The master equation becomes as follows.
\begin{widetext}
	\begin{align}
	\frac{d}{dt} \rho_S (t) = - \frac{4 \pi^2 a^2 \rho_C^2}{\mu_R^2} & \sum_{\vec{l}, \vec{l'}} \sum_{\gamma, \gamma'} \sum_{\alpha, \alpha'} Tr_B \int_0^{\infty} [ \hat{c}^{\dagger}_{ \vec{l}, \alpha } \hat{c}_{ \vec{l}, \alpha' } + \frac{1}{\sqrt{\rho_C V}} \sum_{ \vec{m} } \sum_{ \beta } A_{ \vec{m}, \vec{l} } \left( u_{ \vec{m} } + v_{ \vec{m} } \right) \left( \hat{h}_{ \vec{m}, \beta } + \hat{h}^{\dagger}_{ \vec{m}, \beta } \right) \nonumber \\
	\times & \left( \hat{c}^{\dagger}_{ \vec{l}, \uparrow } \hat{c}_{ \vec{l} + \vec{r}, \uparrow } e^{- i \delta_R t} + \left( \hat{c}^{\dagger}_{ \vec{l}, \uparrow } \hat{c}_{ \vec{l} + \vec{r}, \downarrow } + \hat{c}^{\dagger}_{ \vec{l}, \downarrow } \hat{c}_{ \vec{l} + \vec{r}, \uparrow } + h.c. \right) + h.c. \right), \nonumber \\
	& \times [ \hat{c}^{\dagger}_{ \vec{l'}, \gamma } \hat{c}_{ \vec{l'}, \gamma' } + \frac{1}{\sqrt{\rho_C V}} \sum_{ \vec{m'} } \sum_{ \beta' } A_{ \vec{m'}, \vec{l'} } \left( u_{ \vec{m'} } + v_{ \vec{m'} } \right) \left( \hat{h}_{ \vec{m'}, \beta' } + \hat{h}^{\dagger}_{ \vec{m'}, \beta' } \right) \nonumber \\
	& \times \left( \hat{c}^{\dagger}_{ \vec{l'}, \uparrow } \hat{c}_{ \vec{l'} + \vec{r'}, \uparrow } e^{- i \delta_R (t-s)} + \left( \hat{c}^{\dagger}_{ \vec{l'}, \uparrow } \hat{c}_{ \vec{l'} + \vec{r'}, \downarrow } + \hat{c}^{\dagger}_{ \vec{l'}, \downarrow } \hat{c}_{ \vec{l'} + \vec{r'}, \uparrow } + h.c. \right) + h.c. \right), \rho_S (t) ]]
	\end{align}
\end{widetext}
Using the fermionic anticommutator relations in Eq. (\ref{anticomm1}) and Eq. (\ref{anticomm2}), terms without exponential factors vanish.

Next, due to turbulent fluctuations in the exponential term, the integral vanishes at infinity. Furthermore, we will observe terms with the factor of $e^{\pm 2 i \delta_R t}$, which signify resonance and are dropped by the use of the rotating wave approximation.

By applying further the anticommutator relations in Eq. (\ref{anticomm1}) and Eq. (\ref{anticomm2}), we observe that terms with the patterns of alternating spins i.e. $\hat{c}^{(\dagger)}_{\vec{j}, \uparrow} \hat{c}^{(\dagger)}_{\vec{j'}, \downarrow} \hat{c}^{(\dagger)}_{\vec{k}, \uparrow} \hat{c}^{(\dagger)}_{\vec{k'}, \downarrow}$ and vice versa cancel each other out.

Taking the trace with respect to the bath, terms with bath operators of the same spin i.e. $\hat{h}^{(\dagger)}_{\vec{m}, \alpha} \hat{h}^{(\dagger)}_{\vec{m'}, \alpha}$, where $\alpha \in \{ \uparrow, \downarrow \}$ vanish.

After the mentioned procedure, the resulting master equation is as follows.
\begin{widetext}
	\begin{align}
	\frac{d}{dt} \rho_S (t) = \frac{2 \pi^2 a^2 \rho_C}{i V \delta \mu_R^2} \sum_{\vec{l}, \vec{l'}} \sum_{\vec{m}} & A_{\vec{m}, \vec{l}}^2 \left( u_{\vec{m}} + v_{\vec{m}} \right)^2 \nonumber \\
	\times & \{ - [ \hat{c}_{\vec{l'}, \uparrow}^{\dagger} \hat{c}_{\vec{l'} + \vec{r'}, \uparrow}, [ \hat{c}_{\vec{l}, \downarrow}^{\dagger} \hat{c}_{\vec{l} + \vec{r}, \downarrow}, \rho_S (t) ] ] + [ \hat{c}_{\vec{l'}, \downarrow}^{\dagger} \hat{c}_{\vec{l'} + \vec{r'}, \downarrow}, [ \hat{c}_{\vec{l}, \uparrow}^{\dagger} \hat{c}_{\vec{l} + \vec{r}, \uparrow}, \rho_S (t) ] ] \nonumber \\
	& + [ \hat{c}_{\vec{l'} + \vec{r'}, \downarrow}^{\dagger} \hat{c}_{\vec{l'}, \downarrow}, [ \hat{c}_{\vec{l} + \vec{r}, \uparrow}^{\dagger} \hat{c}_{\vec{l}, \uparrow}, \rho_S (t) ] ] - [ \hat{c}_{\vec{l'} + \vec{r'}, \uparrow}^{\dagger} \hat{c}_{\vec{l'}, \uparrow}, [ \hat{c}_{\vec{l} + \vec{r}, \downarrow}^{\dagger} \hat{c}_{\vec{l}, \downarrow}, \rho_S (t) ] ] \nonumber \\
	& - [ \hat{c}_{\vec{l'}, \uparrow}^{\dagger} \hat{c}_{\vec{l'} + \vec{r'}, \uparrow}, [ \hat{c}_{\vec{l} + \vec{r}, \uparrow}^{\dagger} \hat{c}_{\vec{l}, \uparrow}, \rho_S (t) ] ] + [ \hat{c}_{\vec{l'}, \downarrow}^{\dagger} \hat{c}_{\vec{l'} + \vec{r'}, \downarrow}, [ \hat{c}_{\vec{l} + \vec{r}, \downarrow}^{\dagger} \hat{c}_{\vec{l}, \downarrow}, \rho_S (t) ] ] \nonumber \\
	&  + [ \hat{c}_{\vec{l'} + \vec{r'}, \uparrow}^{\dagger} \hat{c}_{\vec{l'}, \uparrow}, [ \hat{c}_{\vec{l}, \uparrow}^{\dagger} \hat{c}_{\vec{l} + \vec{r}, \uparrow}, \rho_S (t) ] ] - [ \hat{c}_{\vec{l'} + \vec{r'}, \downarrow}^{\dagger} \hat{c}_{\vec{l'}, \downarrow}, [ \hat{c}_{\vec{l}, \downarrow}^{\dagger} \hat{c}_{\vec{l} + \vec{r}, \downarrow}, \rho_S (t) ] ] \}
	\end{align}
\end{widetext}

Using still the fermionic anticommutator relations, we observe that some of these commutators are Hermitian conjugates of another. And thus we obtain the master equation in Eq. (\ref{fermifermimasteqn}).


\end{document}